\documentstyle[epsfig]{aipproc}

\begin{document}
\title{
Bosons and Environment
\footnote {\rm Contributed Paper to 
{\it Spin-Statistics Connection and Commutation Relations}, 
R.C.Hilborn and G.M.Tino eds. (American Institute of
Physics, New York, Sept. 2000)} }

\author{Enrico Celeghini$^*$ and Mario Rasetti$^{\dagger}$}
\address{$^*$Dipartimento di Fisica and Sezione INFN\\
Universit\`a di Firenze, I50125 Firenze, Italy\\
$^{\dagger}$Dipartimento di Fisica and Unit\`a INFM\\
Politecnico di Torino, I10129 Torino, Italy}

\maketitle

\begin{abstract}
The role of the background in bosonic quantum statistics is discussed 
in the frame of a new approach in terms of coherent states. 
Bosons are indeed detected in different physical situations where they
exhibit different and apparently unconnected properties. Besides
Bose gas we consider bosons in particle physics and bosons in harmonic traps. 
In particle physics bosons are dealt with in a context where the 
number of observed particles is finite: here the relevant feature are 
the canonical commutation relations, which we shall show to 
be related to a Boltzmann-like distribution. A further 
case is Bose condensate in harmonic traps, where discrete spectrum 
leads us to predict, below $T_c$, a new critical temperature. The unified 
approach proposed shows that all differences can be ascribed to the environment.  
\end{abstract}

\subsection*{Introduction}

Bosons are discussed in different areas of physics with the implicit 
assumption that none of their properties is related to the characteristics  
of the Fock space where their states are defined (continuum or discrete) or 
to the number of particles (finite or, essentially, infinite).
We discuss here different physical paradigms, all in Fock space and with 
unlimited occupation numbers, showing that such assumption is incorrect: 
different environments are not equivalent and 
unexpected relations with the background can be found.

We introduce first the algebraic approach to quantum statistics recently 
developed, showing that an equivalent but more powerful definition of bosons 
can be stated in terms of coherent states of the non compact algebra $su(1,1)$
\cite{bos}. In the necessary definition of consistent limit procedures for each 
physical problem, the analyticity properties imposed by group theory will be 
shown to play an essential role. 
The applications considered will deal with particle physics (PP), Bose gas, and 
Bose-Einstein condensation (BEC). 
In PP the interest is focused on canonical commutation relations (CCR), the 
energy spectrum is continuum and the number of particles is limited. 
The Bose gas is well exemplified by the blackbody radiation -- once more with 
continuum spectrum but centered on Gibbs and Bose postulates and described in 
terms of finite density (and infinite number) of particles -- and we consider it 
only to compare it with BEC where, together with the postulates, one has discrete 
spectrum and finite number of particles.

It will be shown that in PP bosons satisfy a Boltzmann-like statistics, and just 
because of that they are related to $h(1)$. There follows that they obey CCR, which 
are thus the correct commutation relations in this physical situation (even though, 
as we shall see, not in others).
We shall then argue that the standard approach working for the blackbody, cannot be 
extended to BEC: in the latter for $T < T_c$, finite density approximation fails as 
the spectrum discreteness becomes relevant and a new physics emerges. In particular  
equipartition does not hold any longer and a new critical temperature is found, where 
the specific heat exhibits a spike.

\subsection*{Algebras and Coherent States}

A relation between each statistics and a well defined algebra, has been 
found in 1998 \cite{bos} in terms of coherent states \cite{Per}. 
This has allowed us to select specific raising and lowering operators 
among those that one can arbitrarily introduce into Fock space.
Bose-Einstein definition of bosons is indeed given by requiring the density 
matrix $\rho$ to be a multiple of the identity ${\bf I}$ on the Fock space 
${\cal F}$, with no need of referring to any specific operator.  
One is then free to consider the operators $a^\pm$ relating spaces with 
different number of particles -- ${\cal F}_n$ and ${\cal F}_{n+1}$ -- such 
that 
\begin{eqnarray}
a^+ |n\rangle = f(n+1) |n+1\rangle \; , \label{fn}  
\end{eqnarray}
where $f(n)$ is an arbitrary function. We can assume $f(n) \equiv \sqrt{n}$
but also {\it e.g.} $n$ or $e^n$. 
The choice of $a^\pm$ usually done -- that derives from the Weyl-Heisenberg 
algebra $h(1)$ of position and momentum operators and implies CCR -- is 
arbitrary or, more precisely, appears to be an independent postulate.
In what follows such postulate will be shown, following \cite{bos}, to be 
related to a specific quantum statistics even though, at first sight, an 
unsatisfactory one: only a careful discussion of the limit procedure (taking 
into account correctly the structure of the Fock space) clarifies that for 
bosons in the physical background of PP (and only in it) consistency holds. 

We describe now briefly the procedure that leads from each algebra to 
its characteristic number of states of the system $W\{{\bar n}\}$; more details 
can be found in ref. \cite{bos}.
Let us start from $h(1)$ (often considered as $''$the algebra$\, ''$ of bosons) 
and its coherent states. If we follow Condon-Shortley conventions, we have
\begin{eqnarray}
a^+ |n\rangle \,=\, \sqrt{n+1}\, |n+1\rangle \; , \;\;\;\;\;\;\;\;
e^{a^+}\, |0\rangle \,=\, \sum\, \frac{1}{\sqrt{n!}}\, |n\rangle \; .\nonumber
\end{eqnarray}
If we disregard such conventions and do not fix any gauge, an arbitrary  
phase (possibly a function also of an external parameter $t$, that 
we may assume to describe time) must be considered in front of each addendum 
of the coherent state, and the coefficients in the sum are exactly those needed to 
describe the $1$-mode Boltzmann statistics (see {\it e.g.} \cite{Hua}). Analogously, 
$2$-modes are realized operating on the $2$-mode vacuum $|0,0\rangle$ with 
$\;{a_1}^++{a_2}^+$:
\begin{eqnarray}
e^{{a_1}^+ +\, {a_2}^+}\;\; |0,0\rangle \;\;=\;\; 
\sum \;\frac{{\rm e}^{i \phi(n_1,n_2, t)} 
}{\sqrt{n_1! n_2!}}\; |n_1,n_2\rangle \; .\nonumber
\end{eqnarray}
In order to generalize to $M$-modes we have now only to consider the $M$-mode vacuum
and the operator $\sum_{i=1}^M\,  {a_i}^+$ (in a more formal language, we operate on 
the highest weight of the direct product representation with the iterated coalgebra 
$\Delta^M (a^+) = \sum {a_i}^+$ \cite{Abe}):
\begin{eqnarray}
   \exp \left [ \Delta^M  \left( a^+ \right )\right ] \;\; |0,0,\dots ,0 \rangle
  \;\;=\;\; \sum_{\{{\bar n}\}} \, \, 
    \frac{ {\rm e}^{i \phi({\bar n}, t)} }{\sqrt{n_1 ! \, n_2 ! \dots n_M !}}
 \, |n_1,n_2,\dots ,n_M \rangle \; . 
    \label{cocoh}  
\end{eqnarray}
This is equivalent to
\begin{eqnarray}
W\{{\bar n}\} \;\;\propto\;\;  \prod_{i} 
    \frac{1}{n_i!} \; , \label{ccc}   
\end{eqnarray} 
{\it i.e.} to the definition of Boltzmann statistics \cite{Hua}, that we thus 
obtain from the algebra $h(1)$ -- which contains the canonical commutation relations 
$[\psi(x),\psi^\dagger(y)] = \delta (x-y)$ as the relations $[a_j^-,a_{\ell}^+] = 
\delta_{j\ell}$ written in configuration space. Vice versa eq.(\ref{ccc}) {\it i.e.} 
Boltzmann statistics can be rewritten as eq.(\ref{cocoh}) and thus requires $f(n) = 
\sqrt{n}$ that implies $h(1)$ and its CCR.

We look now for an algebra, if any, related to Bose-Einstein statistics. It turns 
out to be sufficient selecting $f(n) = n$ to obtain 
\begin{eqnarray}
   \exp \left [ \Delta^M  \left( a^+ \right )\right ] \; |0,0,\dots ,0 \rangle
 \;\; =\;\; \sum_{\{{\bar n}\}} \;\;   {\rm e}^{i \phi({\bar n}, t)}   
 \;\; |n_1,n_2,\dots ,n_M \rangle \; . 
    \label{su}  
\end{eqnarray}
Manifestly in this case all states in Fock space have equal probability of being 
occupied. Bose statistics appears thus to be related to $f(n) = n$. With this choice  
eq. (\ref{fn}) leads us to identifying $a^+$ with the raising operator of the 
representation ${\cal D}_{1/2}^{+}$ of $su(1,1)$. Eq.(\ref{su}) describes thus the 
$M$-mode coherent states of the $({\cal D}_{1/2}^{+})^{\otimes M}$ representation 
of $su(1,1)$ (see \cite{bos}).

\subsection*{Canonical Commutation Relations}

The results discussed in the previous section -- surprising as they may sound --
relate on the one hand CCR to the Boltzmann statistics and on the other bosons 
to $su(1,1)$. This apparently uncertain description signals indeed a more subtle 
and complex situation: bosons behave differently in different physical contexts.
To demonstrate such thesis, we start from eq.(\ref{su}) and we follow 
\cite{Hua}, collecting energy levels into cells, each containing $g$ levels. 
If cell $k$, of energy $\epsilon_k$, contains $N_k$ particles, then the number 
of states corresponding to the collection of occupation numbers $\{\bar N\}$ is 
given by 
\begin{eqnarray}
    W\{{\bar N}\} \;\propto \; \prod_{k} 
    \frac{\Gamma (N_k + g)}{\Gamma 
    ( g )\Gamma (N_k+1)} \; .  \label{bos}   
\end{eqnarray} 
Eq.(\ref{bos}), that can be assumed as starting point of any bosons description in 
quantum statistics, is nothing but, as discussed in \cite{bos}, the weight of the 
coherent state corresponding to the representation ${\cal D}_{g/2}^{+}$ of $su(1,1)$. 
By construction, such weight provides the statistics of a finite number of bosons in 
Fock space with discrete spectrum. However, physical situations encountered in 
experiments (except for magnetic traps where BEC is currently investigated) are 
typically different. Appropriate limit procedures must thus
be considered for eq.(\ref{bos}); these procedures, peculiar of each 
problem, must be included in the consistent definition of what bosons are 
in each physical situation. 

In PP the spectrum is continuum ({\it e.g.} in momentum space) while the number 
of particles remains finite: the correct limit procedure on eq.(\ref{bos}) is 
thus given by the limit for $g\to\infty$ at $N_k$ fixed \cite{Abr}:  
 \begin{eqnarray}
    W\{\bar N\} \;\propto \; {\rm lim}_{g\to\infty}\;  \prod_{k} 
    \frac{\Gamma (N_k + g)}{\Gamma 
    ( g )\Gamma (N_k+1)}\; \; =\;\;  \prod_{k}\; 
    \frac{g^{N_k}}{N_k!} \; . \label{bol}   
\end{eqnarray} 
Eq.(\ref{bol}) is exactly equivalent to eq.(\ref{ccc}) (and to
eq.(\ref{cocoh})) from which it can be directly obtained with the same 
procedure of collecting elementary levels in cells that allows to obtain 
eq.(\ref{bos}) from eq.(\ref{su}) \cite{Hua}.
This means that a finite number of bosons, in a continuum Fock space, 
satisfies Boltzmann's statistics and, thus, CCR. This result, which appears to 
contradict common wisdom, is actually consistent with all properties of bosons:
all statistics -- Bose, Boltzmann and Fermi -- give the same distribution in 
these conditions as, in the physics of the continuum, only states with 
occupation numbers limited to zero and one have probabilities different from 
zero.  The only statistical property that survives the limit is indeed the 
symmetry of the wave functions and the related Bose-Einstein correlation 
\cite{Lor} \cite{ALE}.

\subsection*{Bose-Einstein Condensation} 

For the blackbody the limit is different: once more $g\to\infty$ (because 
the configuration space is continuum) but $N_k\to\infty$ as well (as blackbody 
is a theory of gases). Now the limit prescription is that the density remain  
finite: $0 < N_k/g \equiv \delta_k < \infty$. In this way one recovers the usual 
formulas \footnote {\rm One should not forget that the usual 
formalism is affected by a small but conceptually relevant difficulty: as for 
large energy $\delta \approx 0$, the limit prescriptions are not satisfied and 
a not zero probability is predicted for photons of energy greater than that of 
the entire blackbody.}. 

Let us now compare it with Bose-Einstein condensation in harmonic traps.
A detailed technical discussion can be found in ref. \cite{BEC}; we consider 
here only the aspects more closely related with statistics. 
As a finite fraction of particles migrate into the fundamental level, $\delta_0 
\approx \infty$ and the blackbody prescriptions are badly violated.
In the standard approach, an {\sl ad hoc} delta function is introduced in the 
density of states to allow for a macroscopic occupation of the ground state: 
a procedure which is not adequate in that it does not determine the ground state 
occupation. The theory of BEC for $T < T_c$ should be constructed on a different 
basis, namely not in terms of density but of $N$ -- the total number of particles 
-- (finite) in a discrete spectrum. In particular the continuum approximation 
not only does not permit a rigorous treatment of the macroscopic single state 
occupation characteristic of BEC, but is far from the experimental situation, 
where typically $h\nu /k_B \approx 10\, nK$ whereas the estimated temperature $T_c 
\approx \, 10^2\, nK$. 

The solution to the problem can be found coming back to eq.(\ref{bos}),
without any limit. We can thus easily look for the most probable distribution, 
maximizing the logarithm of $W\{{\bar N}\}$ (constrained by the two conditions 
$\sum_i N_i = N$ and $\sum_i \epsilon_i N_i = E$) with respect to all $N_i$'s. 
One obtains in this way the exact equation for $N_i$ and $g$ finite:
\begin{eqnarray}
\psi(N_i+g)-\psi(N_i+1) \;=\; \alpha + \beta \epsilon_i \;\equiv\; h_i \; ,  
\label{psi} 
\end{eqnarray}
where $\alpha =\alpha (g)$ and $\beta =\beta (g)$ are the two Lagrange 
multipliers, clearly depending on the value of $g$, while $\psi$ is 
the digamma function \cite{Abr}. 

For large argument $\psi$ coincides with the logarithm, thus for $N_i \gg 1$ 
and $g\gg 1$ it is easy to obtain the Bose-Einstein distribution we are 
well accustomed to:
\begin{eqnarray}
  \delta = \frac{1}{e^{h}-1} \; ,\label{cont}
\end{eqnarray}
but these limit assumptions do not hold either in PP (where the $N_i$'s 
are finite) nor in harmonic traps (where both $g$ and the $N_i$'s are finite).
As the digamma function has poles for negative integers only, eq.(\ref{psi}) is 
defined almost everywhere in the complex plane and one could conjecture that it 
provides the solution for BEC. 
Unfortunately it is not so: in the presence of a physical 3-dimensional harmonic 
confining potential, levels are not degenerate (as different frequencies cannot 
be exactly equal) {\it i.e.} $g$ should be set equal to 1, and this leads for 
eq.(\ref{psi}) to a perfectly acceptable but void equation (left hand side 
identically zero, implying $\alpha (1) = \beta (1) =0$). The way out is found in 
the analyticity of group theory representations: while in the combinatorial 
approach $g$ is a multiplicity and thus an integer, in theory of representations 
it is simply a parameter that can assume any strictly positive real value.
We can thus find by analytical continuation for $g \to 1$ the equation for not 
degenerate levels we need. $N_i$ results to be the solution of the implicit 
equation 
\begin{eqnarray}
\psi'(N_i+1) = \alpha' + \beta' \epsilon_i \equiv h_i' \; , 
\label{dis} 
\end{eqnarray} 
where $\displaystyle{\alpha' \equiv \lim_{g\to 1} \alpha (g)/(g-1)}$ and 
$\displaystyle{\beta' \equiv \lim_{g\to 1} \beta (g)/(g-1)}$ are the new 
Lagrange multipliers to be determined by the constraints of fixed $N$ and $E$,
and $\psi'$ is the trigamma function \cite{Abr}, derivative of $\psi$. 
Eq.(\ref{dis}) is one-to-one and can be inverted, giving us the correct 
formula for discrete spectrum, to be used instead of eq.(\ref{cont}):
\begin{eqnarray}
N_i = [\psi']^{-1}(h_i') -1 \quad {\rm for} \quad h_i' <  \pi^2/6 
\qquad {\rm and} \qquad  
N_i = 0 \quad {\rm for} \quad h_i'  \geq \pi^2/6 \; ,  \label{inv} 
\end{eqnarray}
$[\psi']^{-1}$ denoting the inverse function of $\psi'$ and $\pi^2/6 \equiv 
\psi' (1)$. 
The discrete description can now be implemented as follows: one assigns first 
the values of $N$ and $E$; from these the two Lagrange multipliers $\alpha' 
= \alpha' (N,E)$, $\beta' = \beta' (N,E)$ are then obtained imposing the 
constraints with $N_i$ given by the defining equation (\ref{inv}); finally, 
inserting $\alpha '(N,E)$ and $\beta '(N,E)$ in (\ref{inv}) itself, one gets 
$N_i \equiv N_i (\alpha' ,\beta' ) = N_i(N,E)$. 

By inspection, the two formulas (\ref{cont}) and (\ref{inv}) exhibit a similar 
structure, because they have a pole with residue 1 for $h=0$ and they go to zero 
for large values of the argument.
As condensation derives from such a behavior, combined with the feature that 
the number of states increases quadratically with energy, both descriptions are 
correct in predicting the collapse of the atoms in the fundamental level, also 
if the continuous approach is actually unable to describe what happens for $T 
< T_c$. The subtlety here is that the very concept of temperature, as defined in 
the theory of gases, cannot be straightforwardly extended to the condensate.

For $E/N$ large ($T>T_c$, classical region) continuum quantum statistics
should predict indeed for the second Lagrange multiplier $\beta' \propto 
{\displaystyle{\left ( E/N \right )}}^{-1}$, while we find $\beta' \propto 
{\displaystyle{\left ( E/N \right )}}^{-4}$. Also for $T$ large the energy 
is therefore not proportional to the inverse Lagrange multiplier ${\beta'}^{-1}$ 
but to the more complex expression $({\pi}^2/6 - \alpha') {\beta'}^{-1}$.
Temperature can then no longer be obtained, as for a gas, by $T = 
\bigl ( k_B \beta' \bigr )^{-1}$: the whole scheme (including the very  
definition of temperature and its relation with energy, {\it i.e.} 
equipartition) must be reconsidered.

In order to obtain the temperature we resort to the more basic notion of 
entropy, adopting its Shannon information theoretical definition: 
\begin{eqnarray}
S = - k_B \, \sum \, p_i \, \ln p_i \; , 
\nonumber 
\end{eqnarray}
where the $p_i$'s are functions of $N$ and $E$: $p_i \equiv N_i(N,E)/N$.  At 
fixed $N$ we thus straightforwardly obtain $S = S(E)$.  Numerical elaborations 
finally give the temperature
\begin{eqnarray}
T \equiv T(E) = \left [\frac{\partial S}{\partial E}\right ]_N^{-1} \; . 
\nonumber 
\end{eqnarray}    
Once more this function is one-to-one, and from $T = T(E)$ we can obtain 
$E = E(T)$, whence the specific heat 
$\displaystyle{C \equiv \frac{1}{N}\, \left [ \frac{\partial E}{\partial T} 
\right ]_N}$ derives by (numerical) differentiation (see Fig. 1).

\begin{figure}[b!] 
\centerline{\epsfig{file=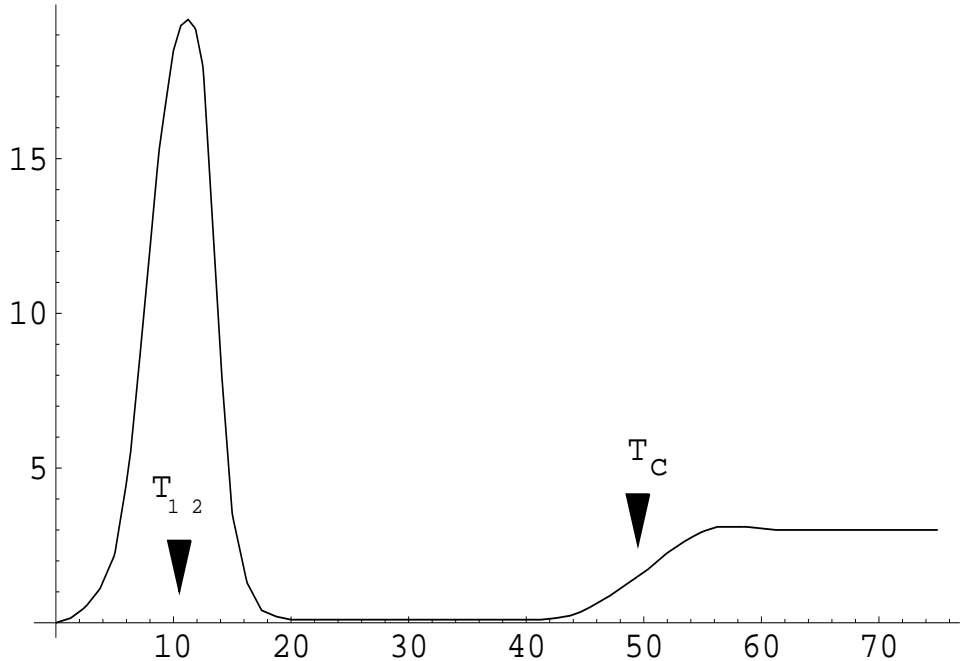,height=3.5in,width=5in}}
\vspace{10pt}
\caption{
 Specific heat $C$ {\sl vs.} $T$ in units $k_B = h \nu =1$ for $N=10^6$ bosons
in an isotropic harmonic trap, as predicted by discrete quantum statistics.}
\label{fig1}
\end{figure}

The result can be checked for $T >T_c$, where $C$ is found to be constant, 
equal to $3 k_B$ (as it should, since the system does not condense and classical 
theory works). 
The physics is completely different for $T < T_c$: $C$ is almost zero 
(as entropy and energy decrease with quite different slopes because only 
a few atoms progressively migrate from the warmer tail of the spectrum into 
the fundamental level) for $T_{1/2} < T < T_c$, and it exhibits a peak for 
$T \approx T_{1/2}$, where condensation becomes a global effect. 
Note that after the spike, below $T_{1/2}$, the specific heat goes to zero 
as $T$ goes to zero, as required by Nernst's theorem.
These results (as well as the complete description of the physics of BEC  
given in \cite{BEC}) are quantitatively correct up to the detailed structure  
of the ground state, and this approach can perhaps be the appropriate background 
in which to introduce a collective perturbation \cite{DGPS}.

In conclusion, the relevance of the algebraic structure of quantum
statistics has been stressed. An important ingredient of the latter 
is analyticity that has been explicitly introduced in the scheme. This 
leads to a unified vision of the problem of bosons. Thanks to such 
approach, apparent differences of bosons in various contexts (particle 
physics, non-interacting gas, condensate) have been shown to be related 
only to the role of the environment. The notion of boson acquires thus 
a primary fundamental role in a unified vision which encompasses different 
definitions and embraces different fields of physics. 
\end{document}